# Design of photonic microcavities in hexagonal boron nitride

Sejeong Kim*, Milos Toth and Igor Aharonovich

Address: School of Mathematical and Physical Sciences, University of Technology Sydney, Ultimo, NSW 2007, Australia

Email: Sejeong Kim – Sejeong.Kim-1@uts.edu.au

## Abstract

*We propose and design photonic crystal cavities (PCCs) in hexagonal boron nitride (hBN) for diverse photonic and quantum applications. Two dimensional (2D) hBN flakes contain quantum emitters which are ultra-bright and photostable at room temperature. To achieve optimal coupling of these emitters to optical resonators, fabrication of cavities from monolithic hBN is therefore required, to maximize the overlap between cavity optical modes and the emitters. Here, we design 2D and 1D PCCs using anisotropic indices of hBN. The influence of underlying substrates and material absorption are investigated, and spontaneous emission rate enhancements are calculated. Our results are promising for future quantum photonic experiments with hBN.*

Hexagonal boron nitride (hBN) has recently emerged as an interesting platform for nanophotonics. This is mainly due to its promising hyperbolic properties [1, 2] as well as the ability to host a range of single photon emitters (SPEs) that are of a great interest for a myriad of nanophotonics and quantum photonic applications [3-10]. However, to further study light matter interaction employing the hBN SPEs, and to realize integrated nanophotonics systems, coupling of the emitters to optical cavities is essential [11-15].

Typically, SPEs can be coupled to optical cavities using two general approaches. The cavity is either made from the material that hosts the emitter (monolithic approach), or the emitter is coupled to a cavity made from a foreign material – also known as a hybrid approach [16]. The former process is preferred when attempting to maximize the field overlap between the emitters and the cavity modes, and is often employed from materials that are amenable to scalable nanofabrication protocols, such as gallium arsenide or silicon [17, 18], and more recently diamond and silicon carbide [19, 20]. The hybrid approach is easier from the fabrication point of view but is inherently limited by the fact that the electric field maxima of optical modes are situated within the cavities, and optimal coupling therefore remains a challenge.

The optical properties of hBN make it an attractive candidate for a monolithic cavity system. In particular, hBN has a wide bandgap of ~ 6 eV which makes it transparent in the visible spectral range that contains the zero phonon lines (ZPLs) of a range of ultra-bright emitters [21]. Furthermore, hBN has properties which are desirable for micro-resonators such as a high chemical stability and an excellent thermal conductivity [22, 23].

In this work, we propose to use hBN for the fabrication of photonic crystal cavities (PCCs). We design two dimensional (2D) PCCs and show that they have high quality (Q) resonances in the visible spectral range, which overlap with the ZPLs of SPEs in hBN [24]. We further optimize the structures and model 1D nanobeam photonic crystals that exhibit a Q-factor in excess of ~ 20,000. In the light of recent progress in direct-write etching of hBN [25], our results are promising for realization of high Q cavities and monolithic coupled systems from this material.

We begin with a 2D photonic crystal that contains a line defect cavity. The L3 cavity has been widely investigated because it was the first to exceed an experimental Q-factor of 10,000 [26, 27]. Birefringence of hBN is accounted for in our study by including both ordinary (nx=ny=1.72) and extraordinary (n=1.84) indices in the 3D finite-difference time-domain (FDTD) method models. The parameters used to define an L3 cavity are shown in Figure 1(a). The cavity consists of a free-standing slab with a triangular photonic lattice with periodicity 'a'. The air hole radius in the mirror region and the radius of two side air holes are fixed at 0.33a and 0.22a respectively. By tuning the periodicity, 'a', one can tune the resonant wavelength whilst preserving the Q-factor. In this study, we used a=270 nm and t=280 nm to place the fundamental mode within the typical emission range of SPEs in hBN (550 nm – 700 nm). Two air holes at the side of the cavity (yellow circles in Figure 1(a)) are reduced in size to minimize radiation losses [28]. For the purpose of design simplicity, we adopt tuning of the side air holes as the only means used to increase the Q-factor. Lastly, the number of photonic crystal layers comprising the cavity is denoted by 'H' which act as a photonic mirror and the thickness of the hBN slab by 't'. The electric field intensity pattern of the L3 cavity calculated using 3D FDTD simulation is shown in Figure 1(b). This is the lowest energy mode in the L3 cavity which is the most widely studied of high-Q 2D photonic cavities.

To optimize the design, we start by increasing the number of photonic crystal layers H, in order to increase the photonic mirror strength and to reduce the in-plane loss of the cavity. As shown in Figure 1(c), the Q-factor of the mode starts to saturate at H ~ 12 because the Q-factor is limited not only by the in-plane component, but also by radiation loss. Considering both the Q-factor and scaling of the simulation time with domain size, we fixed H at 12 for subsequent modeling.

Radiation losses can be minimized by optimizing the side air hole positions, as is shown in Figure 1(d), which reveals that the cavity Q-factor is greatest at a shift distance of 60 nm. Note that the mode resonance red shifts with increasing side hole separation due to an effective increase in cavity length.

Photonic crystal cavities with line defects can be described as Fabry-Perot resonators. Hence, the Q-factor can be enhanced by increasing the cavity length, which can be tuned by varying number of missing air holes. Figure 2(a) shows the Q-factors of L3, L5, L7, and L11 cavities. The electric field intensity profiles of L7 and L11 are also shown in the figure (and that of the L3 cavity is shown in Figure 1(b)). The Q-factor of the fundamental mode increases with effective cavity length. For the L11 cavity, we additionally calculated the effect of the thickness of the 2D photonic crystal slab as is shown in Figure 2(b). The simulated slab thickness is varied up to 300 nm since that is a realistic thickness of typical hBN flakes prepared by the scotch tape exfoliated method. Thicker slabs exhibit stronger light confinement, which results in higher Q-factors.

We also modeled the effect of the refractive index of an underlying substrate, as is shown in Figure 2(c). Because the refractive index of hBN is relatively low compared to that of typical semiconductors, the increase in substrate index greatly degrades the Q-factor of the L11 cavity (and a similar effect is expected for the other cavities as well). This is, however, not a significant problem as the transfer of hBN flakes onto holey substrates is a straightforward process. Furthermore, use of aerogel material that is currently commercially available is another option to achieve low index substrates.

Next we investigate one dimensional nanobeam photonic crystal structures. 1D PCCs are advantageous in that they can have a photonic bandgap between the first and second lowest-lying bands and hence a high Q-factor even when the effective index contrast is low [29]. The combination of a high Q factor and a low refractive index enables a broad range of applications such as flexible photonic crystal devices and high figure of merit sensors [30]. Figure 3(a) shows the electric field intensity profile of the fundamental mode. The cavity is designed by modulating the periodicity whilst fixing the air hole radius [31]. The structure consists of a total of 31 air holes, 15 of which are modulated to create a cavity in the center, and the remaining 8 on each end act as photonic mirrors. In Figure 3(b), we set the periodicity in the mirror region to 260 nm while the air hole radius and the nanobeam width are fixed at 70 nm and 300 nm, respectively. By increasing the thickness of the slab from 200 nm to 300 nm, the Q-factor increases, as is seen in Figure 3(b), and Q-factors in excess of 20,000 can be realized. These values are achieved even without optimization of the remaining structural parameters, and are over an order of magnitude greater than the maximum Q-factor of a low-index 2D cavity. Next, we fix the thickness at 280 nm and tune the structural parameters to further increase the Q-factor of the mode. Figure 2(c) shows the Q-factor plotted as a function of the nanobeam width 'w', showing that the Q-factor has a maximum at a width of 320 nm. Introducing a substrate to the nanobeam degrades the Q-factor as is shown in Figure 2(d). A free-standing structure is preferred, as in the case of the 2D photonic crystals presented earlier.

We also consider the absorption losses in the cavity. Figure 4(a) shows the effect of the absorption by the cavity material. Typical FDTD modeling includes only the real part of the complex refractive index as a structural input which assumes that the material is transparent in the simulated wavelength regime. However, in the case of practical situations in which the cavity material exhibits finite absorption, the imaginary part should also be accounted for. The Q-factor of the fundamental mode in a free-standing nanobeam with a beam-width (w) of 320 nm and a slab thickness (t) of 280 nm is calculated as a function of the imaginary refractive index. An increase in imaginary refractive index (i.e. an increase in material absorption) causes the Q-factor to decrease significantly. Note that realistic Q-factor is determined by $1/Q=1/Q_{ideal}+1/Q_{abs}$ ($Q_{ideal}$: Q-factor with lossless material, $Q_{abs}$: Q-factor with absorption losses) [32], which indicates the material loss restricts the maximum Q-factor that can obtained through experiment. Therefore, including imaginary refractive index for Q-factor calculation provides practical Q-factors.

Finally, we discuss the expected Purcell enhancement due to coupling of emitters to cavity modes. Figure 4(b) shows the spontaneous emission rate ($\gamma_{sp}$) of a quantum emitter in the nanobeam cavity relative to an emitter in free space ($\gamma^0_{sp}$). Because the fundamental mode of the nanobeam has a maximum in the high index region, the Q-factor is calculated across the dashed line (y-axis) shown in the inset (i.e. along the width of the 1D PCC). The dipole emitter in the simulation is y-polarized to match the polarization of the optical mode. The Purcell enhancement has a maximum in the center of the nanobeam where the electric field intensity is greatest. The expected Purcell enhancement is greater than 100 over a range of more than 200 nm along the y-axis, as is seen in Figure 4(b), which relaxes the experimental conditions to precisely position the SPE in the cavity. For a realistic case, with a SPE that exhibits a ZPL at 670 nm that is on resonance with the cavity mode, even moderate Q values of 20,000 (~20 GHz linewidth), will yield a Purcell enhancement of ~ 530.

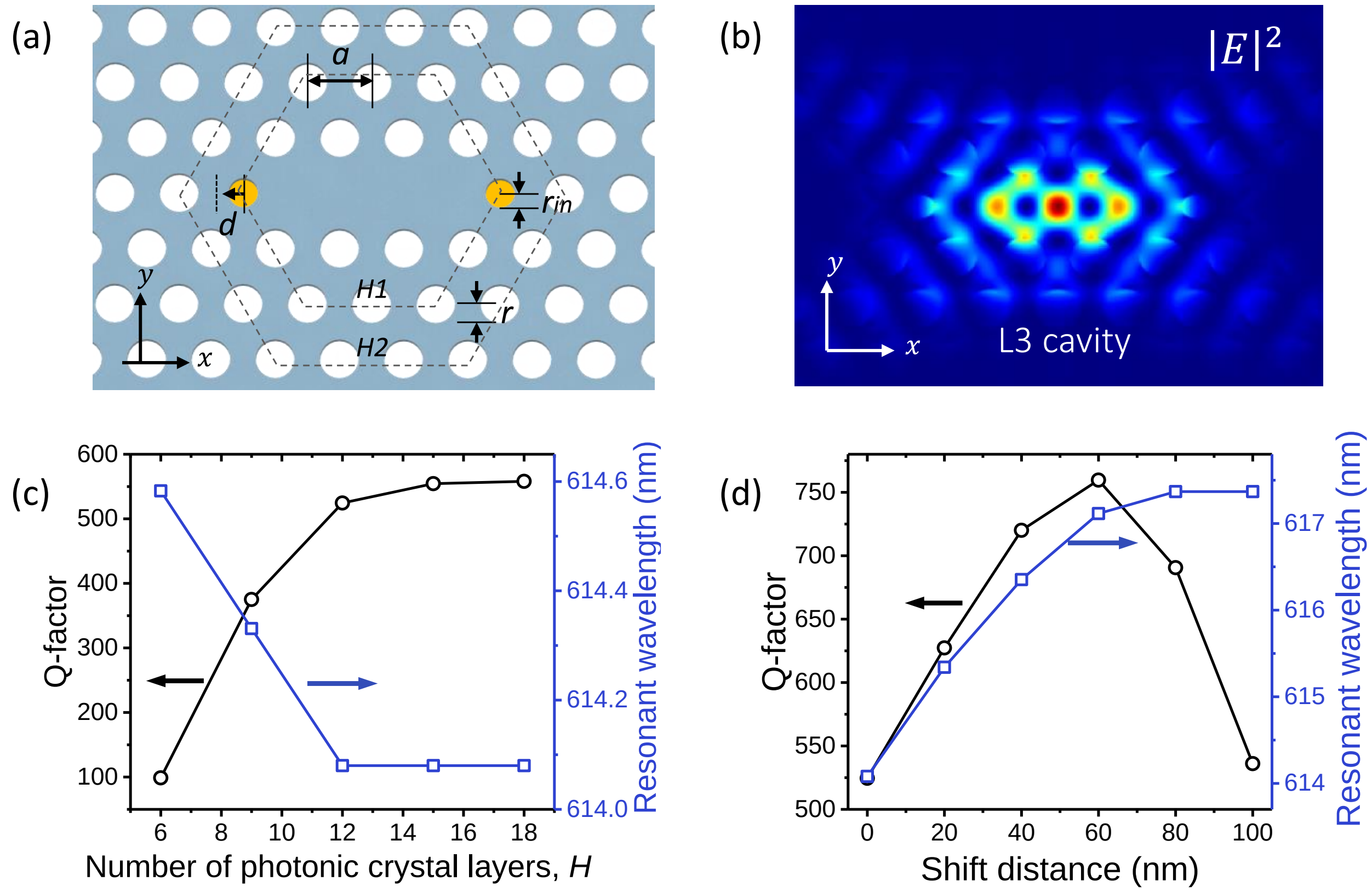

**Figure 1:** (a) Schematic of a 2D photonic crystal with an L3 cavity. The geometric parameters are the following: the period ‘a’, the radius of air holes ‘r’, the radius of two side air holes ‘rin’, the shift distance of the side air holes ‘d’, the thickness of the hBN slab ‘t’ and the number of photonic crystal layers ‘H’. (b) Three-dimensional FDTD simulation of the electric field intensity profile of the fundamental mode of the L3 cavity. Q-factor and the resonant wavelength calculated as a function of (c) the number of photonic crystal layers ‘H’, and (d) the shift distance of the two side air holes ‘d’.

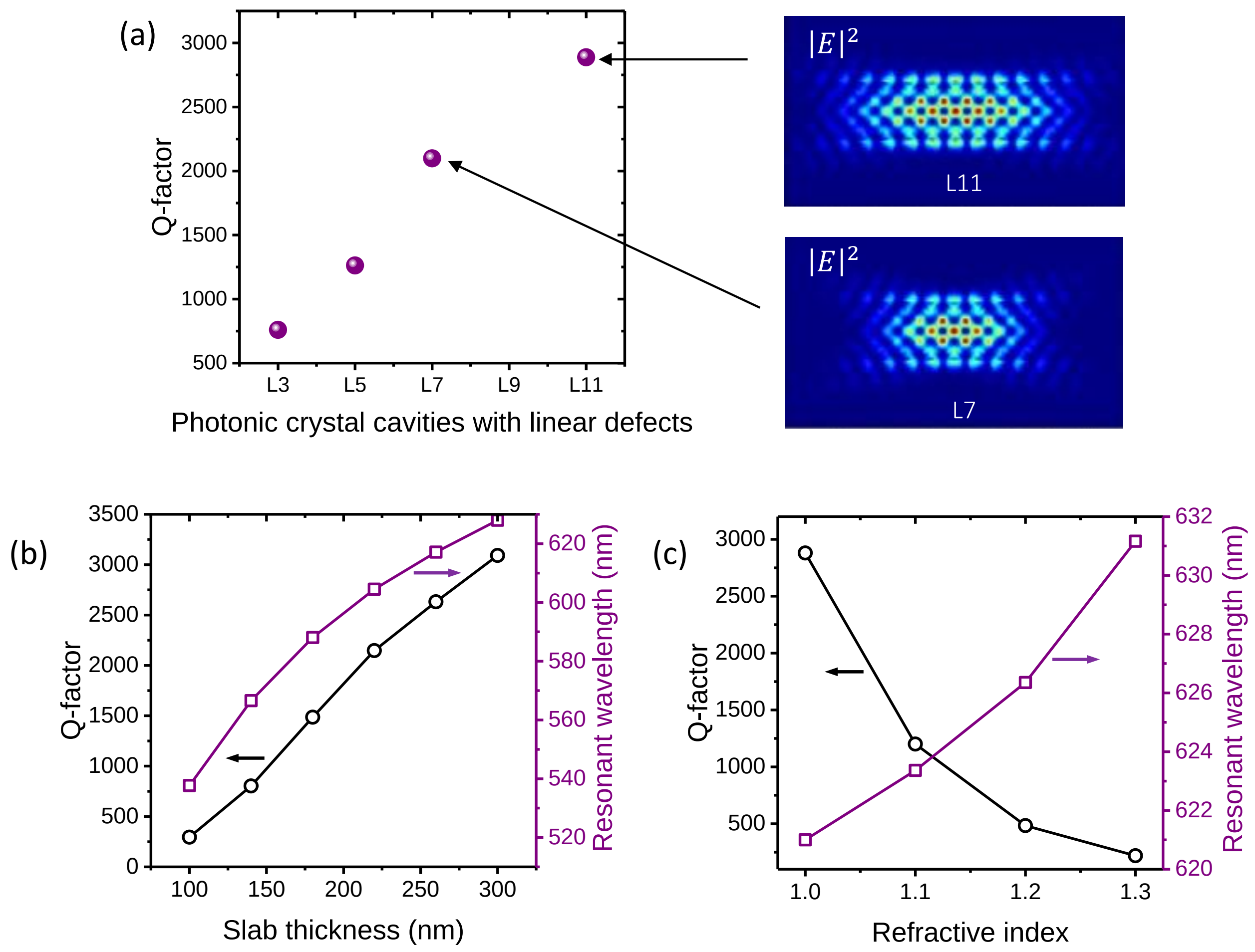

**Figure 2:** (a) Q-factors of various 2D photonic crystal cavities with an increasing number of linear defects, and the electric field intensity profiles of L7 and L11 cavities. Q-factor and resonant wavelength plotted as a function of (b) the slab thickness and (c) the refractive index of an underlying substrate.

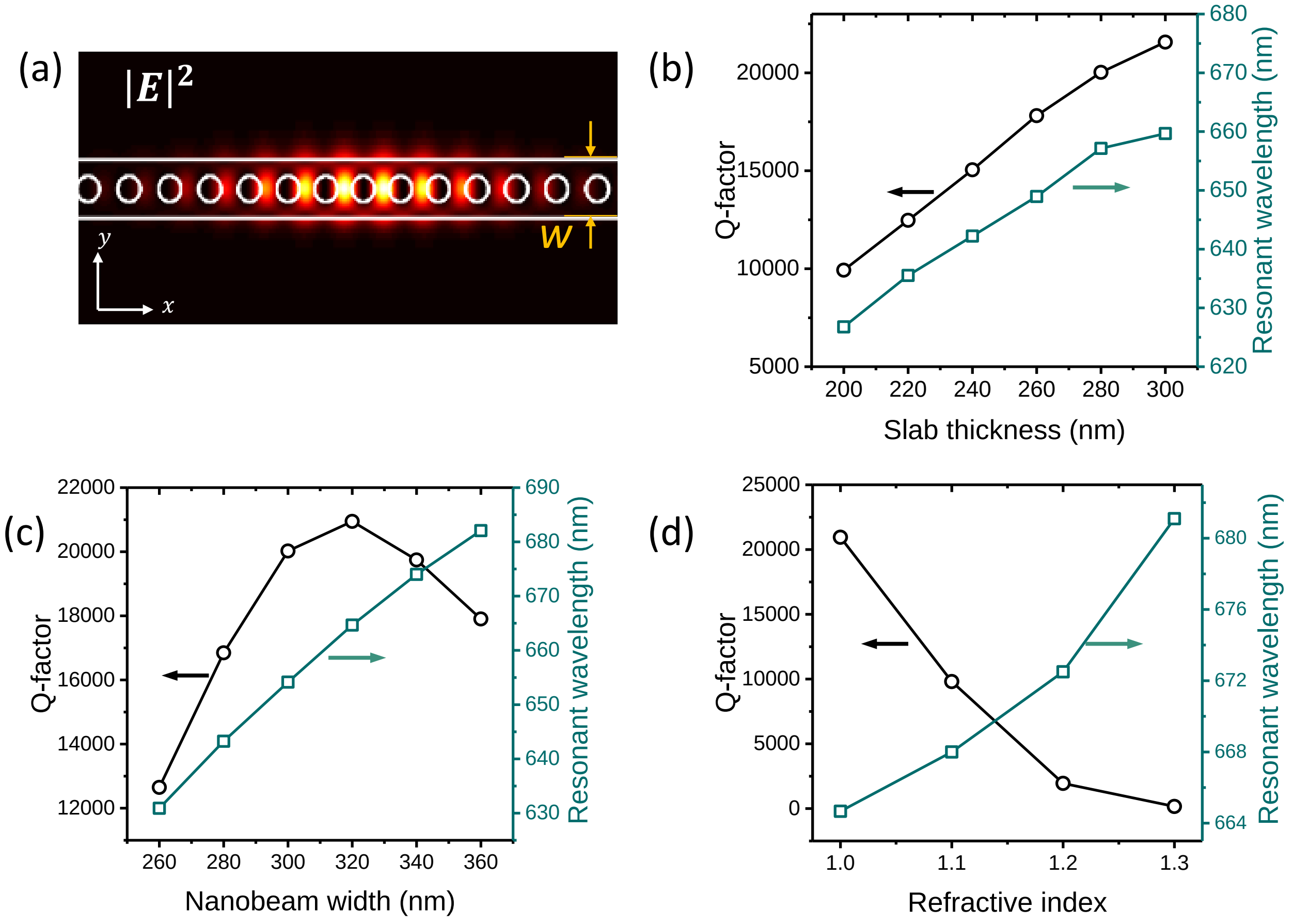

**Figure 3:** (a) Calculated electric field intensity distribution for a 1D photonic crystal cavity. 3D FDTD simulation of Q-factor versus (b) slab thickness, (c) nanobeam width ‘w’, and (d) refractive index of the substrate, respectively.

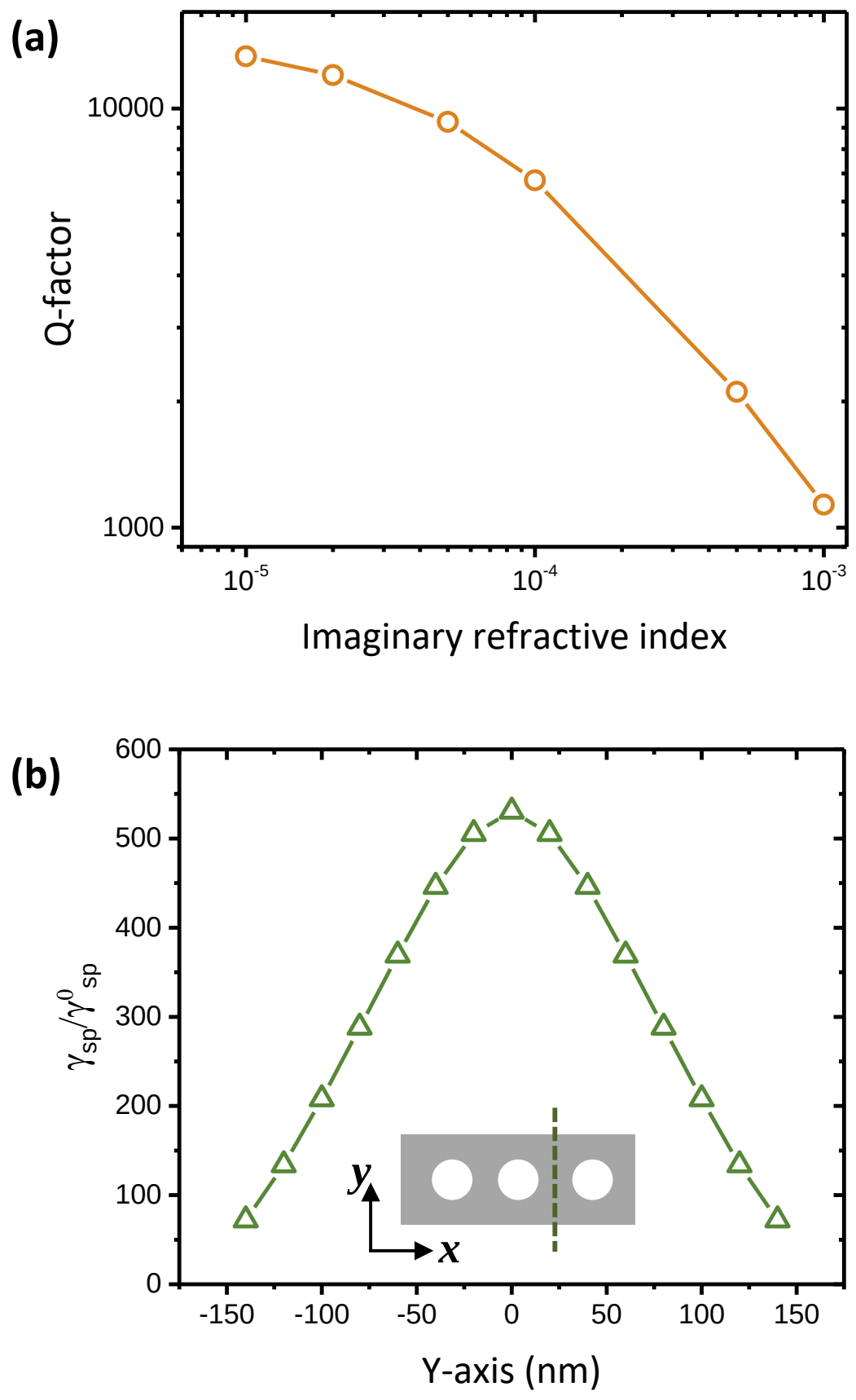

**Figure 4:** (a) Q-factor of the fundamental mode in a 1D photonic crystal cavity versus the imaginary refractive index coefficient of the material. (b) Spontaneous emission rate enhancement of a dipole emitter plotted as a function of distance across the 1D photonic crystal nanobeam cavity.

In summary, we described and optimized a number of 2D and 1D PCC designs in free-standing and supported hBN layers. Linear defect cavities were studied as representatives of 2D photonic crystals. Period modulation of a 1D nanobeam was used to achieve a theoretical Q-factor in excess of 20,000 simply by modulating the beam width of the structure. The effect of the imaginary refractive index on the Q-factor of a nanobeam was simulated, as was the Purcell factor, showing a strong interaction between a dipole emitter and the optical mode. The designs and analyses of the hBN photonic cavities presented in this work will pave a way to a broad range of applications enabled by integrated photonic circuits based on 2D materials.

## Acknowledgements

We thank Prof. Joshua D. Caldwell for useful discussions on hBN refractive index. Financial support from the Australian Research Council (DE130100592, DP140102721), the Asian Office of Aerospace Research and Development grant FA2386-15-1-4044 are gratefully acknowledged.